\documentclass[
 reprint,
 aps,
 prapplied,
 amsmath,amssymb,
 floatfix,superscriptaddress
]{revtex4-2}

\usepackage{graphicx}% Include figure files
\usepackage{dcolumn}% Align table columns on decimal point
\usepackage{bm}% bold math
\newcommand{\Tr}{\operatorname{Tr}}
\usepackage{ulem}
\usepackage{xcolor}
\usepackage{hyperref}

\begin{document}

\preprint{APS/123-QED}

\title{Generation of Schr\"odinger cat-like states via degenerate dual pump spontaneous four-wave mixing in a $\chi^{(3)}$ microring resonator}

\author{Ranjit Singh}
\email{ranjit.singh@mail.ru}
\affiliation{Independent Researcher, Domodedovo, 142000, Moscow region, Russia}

\author{Alexander E. Teretenkov}
\affiliation{Department of Mathematical Methods for Quantum Technologies,
Steklov Mathematical Institute of Russian Academy of Sciences,
8 Gubkina St., Moscow, 119991, Russia}

\date{\today}% It is always \today, today,

\begin{abstract}
We theoretically investigate the generation of non-Gaussian quantum states, specifically Schr\"odinger cat-like states (SCLSs), via degenerate dual-pump spontaneous four-wave mixing in a $\chi^{(3)}$-based microring resonator. By introducing a unitary transformation that exactly decouples the self-phase modulation (SPM) and cross-phase modulation (XPM) terms, we reduce the full nonlinear Hamiltonian to an effective three-mode interaction. The resulting dynamics (decoupled and full Hamiltonians) are studied using the Lindblad master equation, accounting for cavity losses. Unlike semiclassical or parametric approximations, our full quantum mechanical approach explicitly includes quantum pump depletion, which enables the emergence and observation of non-Gaussian features. We compute the Wigner function, photon number distributions, quadrature variances, Fano factor, Schmidt number, and fidelity to characterize the generated states. For the non-dissipative case, we find that the signal mode $\hat{b}_3$ or $\hat{a}_3$ exhibits clear non-Gaussian features with a structured Wigner function and even-dominated photon number distribution, characteristic of an even coherent state. In the presence of dissipation ($\gamma_j = 0.2$), the interference fringes become faint, odd photon numbers appear, and the fidelity with the ideal state remains high ($>0.9$), indicating robustness. The pump mode $\hat{b}_1$ or $\hat{a}_1$ remains Gaussian, while both modes display super-Poissonian statistics and entanglement ($>2$). Our results demonstrate that degenerate dual-pump spontaneous four-wave mixing in microring resonators is a promising platform for generating and controlling cat-like states under dissipative conditions.
\end{abstract}

%\keywords{Schrödinger cat states, second harmonic generation, Wigner quasiprobability distribution, classical simulation.}%Use showkeys class option if keyword
\maketitle

%\tableofcontents

\section{Introduction}
Non-Gaussian quantum states, particularly Schrödinger cat-like states (SCLSs) \cite{Cochrane1999,Ourjoumtsev2006,Sychev2017,SinghTeretenkov2026,SinghTeretenkovMasalov2025,SinghBarinovAmosovMasalov2026}, are essential resources for continuous-variable quantum information processing and quantum sensing, and tests of quantum foundations \cite{Cochrane1999,Ourjoumtsev2006,Delaubert2006}. These states, which are superpositions of macroscopically distinct coherent states, exhibit strong non-classical features such as negative Wigner functions and quadrature squeezing. Generating such states in an integrated, scalable platform remains a key challenge.

Optical microresonators with third-order nonlinearity ($\chi^{(3)}$) offer a compact and power-efficient environment for quantum state engineering. Spontaneous four-wave mixing (FWM) is a well-established process for producing squeezed light and entangled photon pairs. In particular, degenerate dual-pump spontaneous FWM (DP-SFWM) \cite{Vorobyev2025, Ulanov2025,Zhang2021,Wu2022,Vernon2019}, where two pump photons at frequencies $\omega_1$ and $\omega_2$ annihilate to create two degenerate photons at $\omega_3 = (\omega_1+\omega_2)/2$, has been shown to generate quadrature squeezing  under suitable conditions \cite{Ulanov2025}. However, practical implementations must contend with parasitic nonlinear effects: Bragg scattering (BS-FWM), self-phase modulation (SPM), cross-phase modulation (XPM), and single-pump FWM (SP-SFWM) processes that can degrade the quantum state. While SP-SFWM can be suppressed by dispersion engineering, SPM and XPM induce unwanted frequency shifts that require compensation \cite{Ulanov2025}.

In this work, we present a theoretical study of SCLS generation via DP-SFWM in a $\chi^{(3)}$ microring resonator. We start from a full Hamiltonian that includes FWM, SPM, and XPM. By applying a unitary transformation that exploits the total photon number conservation, we show that the SPM and XPM terms can be exactly decoupled provided certain relations among the nonlinear coupling coefficients hold. This leaves a clean three-mode interaction Hamiltonian amenable to analytic and numerical treatment. We then consider the dynamics generated by the decoupled and full Hamiltonian and solve the Lindblad master equation for both the non-dissipative and dissipative regimes using QuTiP \cite{johansson2012qutip,JOHANSSON20131234}, focusing on the dynamics of the pump mode $\hat{b}_1$ (in decoupled regime) or $\hat{a}_1$ (in full or coupled regime) and the signal mode $\hat{b}_3$ (in decoupled regime) or $\hat{a}_3$ (in full or coupled regime). We characterize the generated states via Wigner functions, photon number distributions, quadrature variances, Fano factor, Schmidt number, and fidelity.

Our main findings are: (i) At the extremal interaction time $\tau = 0.190$, the signal mode reaches a non-Gaussian state with a two-peak structure in its Wigner function and an even-photon-number distribution, hallmark features of an even coherent state. (ii) The pump mode remains Gaussian. (iii) Dissipation ($\gamma_j = 0.2$) partially washes out the interference fringes and introduces odd photon numbers, but the state retains high fidelity ($>0.9$) with the ideal non-dissipative state. (iv) Both modes exhibit super-Poissonian statistics and the bipartite system shows entanglement with Schmidt number $>2$. These results establish DP-SFWM in microring resonators as a robust method for on-chip generation of cat-like states.

\section{Hamiltonian and decoupling of nonlinear frequency shift}
Consider three optical modes $\hat{a}_1$, $\hat{a}_2$, and $\hat{a}_3$ with frequencies $\omega_1$, $\omega_2$, and $\omega_3$ in a $\chi^{(3)}$ microring resonator. Parasitic processes such as SP-SFWM and BS-FWM can be suppressed, e.g., via dispersion engineering that splits \cite{Ulanov2025} the relevant frequency components (e.g., $X_{-2}$ and $X_{+2}$). We therefore consider only the dominant contributions: the target dual-pump four-wave mixing (FWM) interaction, self-phase modulation (SPM), and cross-phase modulation (XPM). The corresponding interaction Hamiltonian \cite{Vaidya2020} reads
\begin{align}
\hat{H}_{\text{int1}} &= \hat{H}_{\text{FWM}} + \hat{H}_{\text{SPM}} + \hat{H}_{\text{XPM}}, \\
\hat{H}_{\text{FWM}} &= \hbar g \bigl( \hat{a}_1 \hat{a}_2 \hat{a}_3^{\dagger 2} + \hat{a}_1^\dagger \hat{a}_2^\dagger \hat{a}_3^2 \bigr), \\
\hat{H}_{\text{SPM}} &= \hbar \sum_{j=1}^{3} g_j \hat{a}_j^{\dagger 2} \hat{a}_j^2, \\
\hat{H}_{\text{XPM}} &= \hbar \sum_{1 \le i < j \le 3} g_{ij} \hat{a}_i^\dagger \hat{a}_i \hat{a}_j^\dagger \hat{a}_j.
\label{eq:hamiltonian}
\end{align}
	
	The coupling constant $g = |g|e^{i\Delta\omega t}$ contains the linear phase mismatch 
	$\Delta\omega = 2\omega_3 - \omega_1 - \omega_2$, which arises from material dispersion 
	in the $\chi^{(3)}$ medium. Efficient four-wave mixing requires phase matching, i.e., 
	$\Delta\omega = 0$. Here we assume this condition is satisfied via dispersion 
	engineering, so that $g = |g|$ becomes a time-independent constant. For simplicity 
	we set $\hbar = 1$.
	
	\subsection{Heisenberg Equations of Motion}
	
	The Heisenberg equation for an operator $\hat{a}_j$ is:
\begin{align}
		\frac{d}{dt} \hat{a}_1 &= -ig \hat{a}_2^{\dagger} \hat{a}_3^{2} -i \hat{\Omega}_1) \hat{a}_1, \label{eq:h1} \\
		\frac{d}{dt} \hat{a}_2 &= -ig \hat{a}_1^{\dagger} \hat{a}_3^{2} -i \hat{\Omega}_2 \hat{a}_2, \label{eq:h2} \\
		        \frac{d}{dt} \hat{a}_3 &= -i2g^{*} \hat{a}_1 \hat{a}_2 \hat{a}_3^{\dagger} -i  \hat{\Omega}_3 \hat{a}_3. \label{eq:h3}
	\end{align}
  	
	The nonlinear frequency shifts are given by the Hermitian operators:
	
	\begin{equation}
		\hat{\Omega}_i = \sum_{j=1}^3 M_{ij} \hat{n}_j, \quad 
		M = 
		\begin{pmatrix}
			2g_1 & g_{12} & g_{13} \\
			g_{12} & 2g_2 & g_{23} \\
			g_{13} & g_{23} & 2g_3
		\end{pmatrix}.
		\label{eq:delta_omega_matrix}
	\end{equation}
	
	Explicitly:
	
	\begin{align}
		\hat{\Omega}_1 &= 2g_1 \hat{n}_1 + g_{12} \hat{n}_2 + g_{13} \hat{n}_3, \label{eq:dw1}\\
		\hat{\Omega}_2 &= g_{12} \hat{n}_1 + 2g_2 \hat{n}_2 + g_{23} \hat{n}_3, \label{eq:dw2}\\
		\hat{\Omega}_3 &= g_{13} \hat{n}_1 + g_{23} \hat{n}_2 + 2g_3 \hat{n}_3. \label{eq:dw3}
	\end{align}
	where $\hat{n}_j = \hat{a}^{\dagger}_j\hat{a}_j$.
	
\subsection{Exact Decoupling of Self- and Cross-Phase Modulation in Four-Wave Mixing}

	The transformation $\hat{a}_j = e^{-i g \hat{N} t} \hat{b}_j$ is introduced to remove the SPM and XPM terms. Using the identities $e^{+i g \hat{N} t} \hat{b}_j e^{-i g \hat{N} t} = \hat{b}_j e^{- i g t}$ to simplify products then yields simple equations for $\hat{b}_j$ that contain only the FWM interaction, leading directly to the final result for all three modes:
	
\begin{align}
    \frac{d\hat{b}_1}{dt} &= -i g \hat{b}_2^\dagger \hat{b}_3^2, \label{eq:db1} \\
    \frac{d\hat{b}_2}{dt} &= -i g \hat{b}_1^\dagger \hat{b}_3^2, \label{eq:db2} \\
    \frac{d\hat{b}_3}{dt} &= -i 2g \hat{b}_1 \hat{b}_2 \hat{b}_3^\dagger, \label{eq:db3}
\end{align}
where $\hat{N} = \sum_{j=1}^3 \hat{n}_j$. The set of equations \eqref{eq:db1}--\eqref{eq:db3} is valid if $\hat{N}$ is a constant of motion, i.e., $[\hat{H}_{\text{int1}}, \hat{N}] = 0$ and $\frac{d\hat{N}}{dt} = 0$, and the nonlinear coupling coefficients satisfy
\begin{equation}\label{eq:coeff123}
\begin{aligned}
    2g_1 + g_{12} &= 2g_{13}, \\
    2g_2 + g_{12} &= 2g_{23}, \\
    g_{13} + g_{23} &= 4g_3,
\end{aligned}
\end{equation}
while the individual couplings satisfy
\begin{equation}\label{eq:coeff4}
g_1 = g_2 = g_3 = \frac{g}{2}, \qquad
g_{12} = g_{13} = g_{23} = g.
\end{equation}
A related compensation of nonlinear frequency shifts was considered experimentally in Ref.~\cite{Ulanov2025}.

\subsection{Interaction Hamiltonian for FWM process}
By applying the transformation $\hat{a}_j = e^{-i g \hat{N} t}  \, \hat{b}_j$ to \eqref{eq:h1}--\eqref{eq:h3} one arrives at \eqref{eq:db1}--\eqref{eq:db3} and interaction Hamiltonian for FWM can be given by
\begin{equation}
    \hat{H}_{\text{int2}} = g\left( \hat{b}_1 \hat{b}_2 \hat{b}_3^{\dagger}\hat{b}_3^{\dagger} + \hat{b}_1^{\dagger}\hat{b}_2^{\dagger}\hat{b}_3\hat{b}_3 \right).
\end{equation}

We note that the operators $\hat{b}_j$ are introduced via an operator-valued phase transformation and do not correspond directly to physical field modes. In contrast, the experimentally accessible observables are associated with the original modes $\hat{a}_j$. While the transformation is convenient for simplifying the dynamics (in the unitary case), it represents a change of representation rather than a physically implemented operation, and care must be taken when interpreting phase-sensitive quantities in terms of measurable observables. In particular, if the state were restricted to a subspace with a fixed total photon number, the transformation would reduce to a c-number phase factor and could be effectively compensated, for instance, by an appropriate choice of the local oscillator phase or frequency in homodyne detection. However, in the present setting the states involve superpositions of different photon-number sectors, and the operator-valued phase cannot be eliminated in this way, which makes its physical interpretation more subtle.

\section{Dissipative dynamics}
The Lindblad master equation for the density matrix $\hat{\rho}$ \cite{BP2002} describing the non-dissipative $(\gamma_j=0)$ and dissipative $(\gamma_j>0)$ dynamics of the three-mode system is
\begin{equation}
    \frac{d\hat{\rho}}{d\tau} = -i[\hat{H}_{\text{int1}}/g,\hat{\rho}] + \sum_{j=1}^{3} \left( \hat{C}_j \hat{\rho} \hat{C}_j^{\dagger} - \frac{1}{2}\{\hat{C}_j^{\dagger}\hat{C}_j, \hat{\rho}\} \right),
    \label{eq:lindblad}
\end{equation}
where $\tau = g t$ is the dimensionless interaction time.

The Lindblad operators $\hat{C}_j$ describe cavity decay of mode $j$:
\begin{align*}
    \hat{C}_j = \sqrt{\gamma_j}\;\hat{a}_j, \qquad j = 1,2,3,
\end{align*} 
with $\gamma_j \ge 0$ the respective dimensionless damping rates (i.e. dimensional  damping rates in the units of inverse $t$ are $g \gamma_j$ in our notation).

We solve equation~\ref{eq:lindblad} numerically with QuTiP for two scenarios: 
(1) the non-dissipative case (\(\gamma_j = 0\)) for \(\hat{H}_{\text{int2}}\); 
(2) both the non-dissipative (\(\gamma_j = 0\)) and the dissipative (\(\gamma_j = 0.2\)) cases for \(\hat{H}_{\text{int1}}\). 
The initial condition is the same for both scenarios, i.e., the density matrix \(\hat{\rho}(0) = |\psi_0\rangle\langle\psi_0|\), where the pure state is chosen as
\[
|\psi_0\rangle = |\alpha_{10}\rangle_1 \otimes |\alpha_{20}\rangle_2 \otimes |0\rangle_3.
\]
At \(\tau=0\), the pump modes \(\hat{a}_1\) and \(\hat{a}_2\) start in coherent states \(|\alpha_{10}\rangle_1\), \(|\alpha_{20}\rangle_2\), and mode \(\hat{a}_3\) starts in the vacuum \(|0\rangle_3\), with an average photon number \(|\alpha_{10}|^2 = |\alpha_{20}|^2 = 9\) and a phase \(\varphi_{10} = \varphi_{20} = \pi/4\).

\section{Generation of Schr\"odinger cat-like states (SCLSs) in the signal mode (\(\hat{b}_3\))}

The quantum statistical properties of the optical modes can be investigated by means of the Wigner quasiprobability distribution \cite{Walls2008,Agarwal2013}. The Wigner function's phase-space portraits reveal subtle features, including (i) interference patterns characteristic of the wave nature of superpositions of macroscopically distinct states (e.g., SCLSs), and (ii) non-Gaussian behavior manifested as negative values of the distribution. Because the dynamics of modes $\hat{b}_1$ and $\hat{b}_2$ are identical, we present results only for $\hat{b}_1$ (representing the pump) and $\hat{b}_3$ (the signal mode). The Wigner functions for $\hat{b}_1$ and $\hat{b}_3$ are then computed using

\begin{equation}
	W_j(x,p) = \frac{1}{2\pi} \int_{-\infty}^{\infty} \langle x - \tfrac{y}{2} | \rho_j | x + \tfrac{y}{2} \rangle e^{ipy} \, dy.\label{eq:wigf}
\end{equation}
% Figure 1: Non-dissipative, mode b1
\begin{figure}[htbp]
    \centering
    \includegraphics[width=0.8\linewidth]{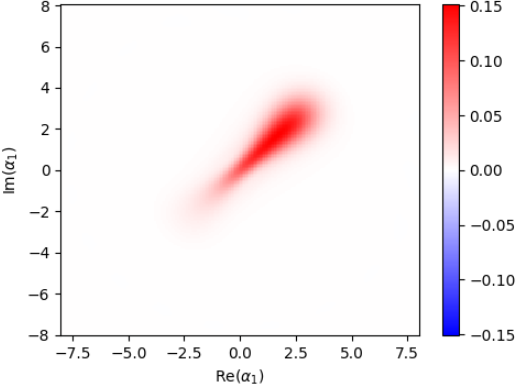}
    \caption{Wigner function for mode \(\hat{b}_1\) under non‑dissipative conditions (\(\gamma_j = 0\)) at \(\tau = 0.190\).}
    \label{fig:fig1nd}
\end{figure}

% Figure 3: Non-dissipative, mode b3
\begin{figure}[htbp]
    \centering
    \includegraphics[width=0.8\linewidth]{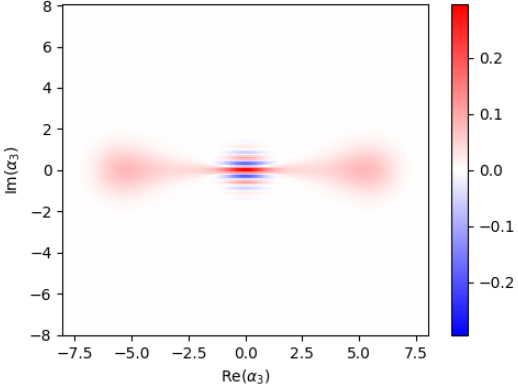}
    \caption{Wigner function for mode \(\hat{b}_3\) under non‑dissipative conditions (\(\gamma_j = 0\)) at \(\tau = 0.190\). A non-Gaussian state is formed. The interference pattern is characteristic of an SCLS. The SCLS is formed at the maximum value of $n_3$ at $\tau=0.190$.}
    \label{fig:fig3nd}
\end{figure}

Figures~\ref{fig:fig1nd} and~\ref{fig:fig3nd} show the Wigner functions (Eq.~\ref{eq:wigf}) for the non‑dissipative case. Figure~\ref{fig:fig3nd} illustrates the generation of a Schr\"odinger cat-like state (SCLS) in the signal mode \(\hat{b}_3\), while the pump mode \(\hat{b}_1\) remains in a classical state (positive Wigner function).

\section{Key statistical properties for applications}

From the Wigner function we extract all statistical moments contributing to the cat-like states and obtain qualitative phase-space portraits (see Figs.~\ref{fig:fig1nd}--\ref{fig:fig3nd}). For a quantitative assessment of relevant quantum statistical observables—such as the squeezing level (quadrature variances and the Fano factor), photon number distributions, and entanglement measures—we perform the corresponding calculations.

\subsection{Mean photon numbers and quadrature variances}

The mean number of photons in each mode \(\hat{b}_j\) is calculated using
\begin{eqnarray}
    n_j (\tau) = \Tr [\hat{n}_j \hat{\rho}_j(\tau)]. \label{eq:nj}
\end{eqnarray}

The variances of the quadrature components are obtained from
\begin{eqnarray}
    \operatorname{Var} (x_j) = \Tr [\hat{x}_j^2 \hat{\rho}_j(\tau)] - \left(\Tr [\hat{x}_j \hat{\rho}_j(\tau)]\right)^2, \label{eq:quad1}\\
    \operatorname{Var}(p_j) = \Tr [\hat{p}_j^2 \hat{\rho}_j(\tau)] - \left(\Tr [\hat{p}_j \hat{\rho}_j(\tau)]\right)^2, \label{eq:quad2}
\end{eqnarray}
where \(\hat{x}_j = (\hat{b}_j + \hat{b}^{\dagger}_j)/\sqrt{2}\) and \(\hat{p}_j = -i(\hat{b}_j - \hat{b}^{\dagger}_j)/\sqrt{2}\) are the quadrature components of mode \(\hat{b}_j\).

\begin{figure}[htbp]
    \centering
    \includegraphics[width=\linewidth]{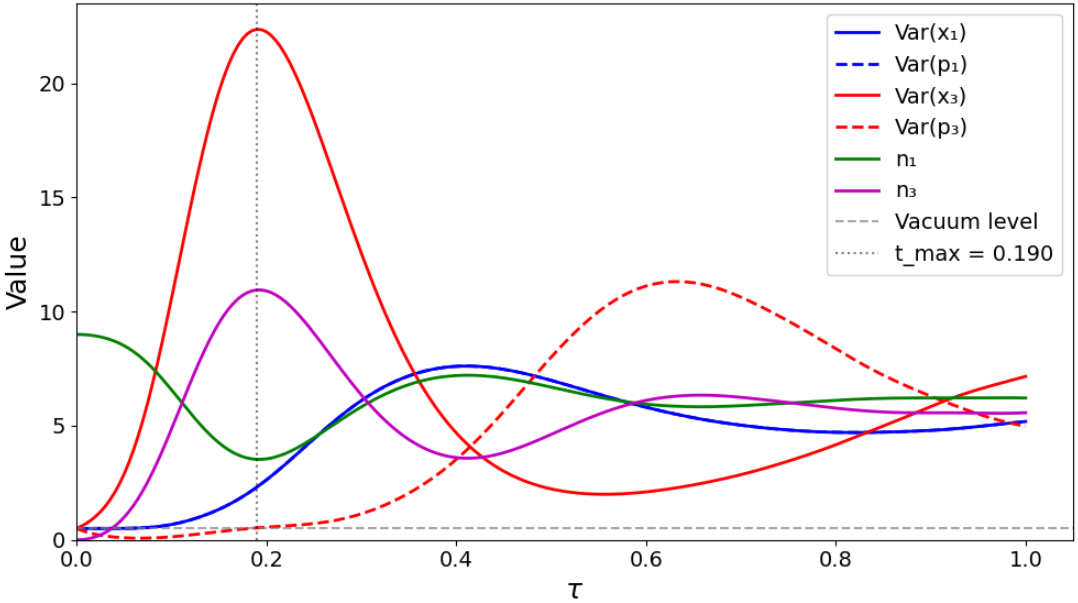}
    \caption{Evolution of (a) the mean photon numbers $n_j$ and (b) the quadrature variances for the FWM process (non‑dissipative).}
    \label{fig:fig1}
\end{figure}

Fig.~\ref{fig:fig1} shows the evolution of (i) the mean photon numbers (Eq.~\ref{eq:nj}) and (ii) the quadrature variances (Eqs.~\ref{eq:quad1} and~\ref{eq:quad2}) for the non‑dissipative FWM process. Schrödinger cat‑like states (SCLSs) form at the first extremal values of the photon numbers: the minimum for mode \(\hat{b}_1\) and the maximum for mode \(\hat{b}_3\), both occurring at \(\tau = 0.190\). At this point, the quadrature variances of mode \(\hat{b}_1\) are \(\operatorname{Var}(x_1) = 2.312092\) and \(\operatorname{Var}(p_1) = 2.312092\), while for mode \(\hat{b}_3\) they are \(\operatorname{Var}(x_3) =  22.362901\) and \(\operatorname{Var}(p_3) = 0.526722\) respectively.

\subsection{Entanglement property of the generated states}
To quantify the degree of entanglement \cite{Ulanov2025} between the two modes \((\hat{b}_1, \hat{b}_2)\) and \(\hat{b}_3\), we employ the Schmidt number \cite{Law2004}, given by
\begin{eqnarray}
    K(\tau) = \frac{1}{\Tr[(\hat{\rho}_{12}(\tau))^2]} = \frac{1}{\Tr[(\hat{\rho}_3(\tau))^2]}, \label{eq:ent}
\end{eqnarray}
with \(\hat{\rho}_{12}\) and \(\hat{\rho}_3\) being the reduced density matrices of the respective modes. For a pure bipartite system, \(K > 1\) signals non‑separability (entanglement). At the extremal point \(\tau = 0.190\) in the non‑dissipative scenario, the Schmidt number evaluates to \(K = 2.0248\), thereby certifying the existence of entanglement.

\subsection{Super-Poissonian statistics}
The nature of the photon statistics—sub‑Poissonian, Poissonian, or super‑Poissonian—can be assessed using the Fano factor. It is defined as
\begin{eqnarray}
    \mathrm{FF}_j(\tau) = \frac{\langle (\hat{n}_j)^2 \rangle - (n_j)^2}{n_j}, \label{eq:ff}
\end{eqnarray}
where the expectation values are taken with respect to \(\hat{\rho}_j(\tau)\). The criterion is: \(\mathrm{FF}_j < 1\) for sub‑Poissonian, \(\mathrm{FF}_j = 1\) for Poissonian, and \(\mathrm{FF}_j > 1\) for super‑Poissonian behavior. At \(\tau = 0.190\), the computed Fano factors (from Eq.~\eqref{eq:ff}) are \(\mathrm{FF}_1 = 3.5071\) (non‑dissipative) for mode \(\hat{b}_1\), and \(\mathrm{FF}_3 = 3.6268\) (non‑dissipative) for mode \(\hat{b}_3\). All these values lie above unity, confirming that the statistics remain super‑Poissonian.

\subsection{Photon number distributions}
\begin{figure}[htbp]
    \centering
    \includegraphics[width=1\linewidth]{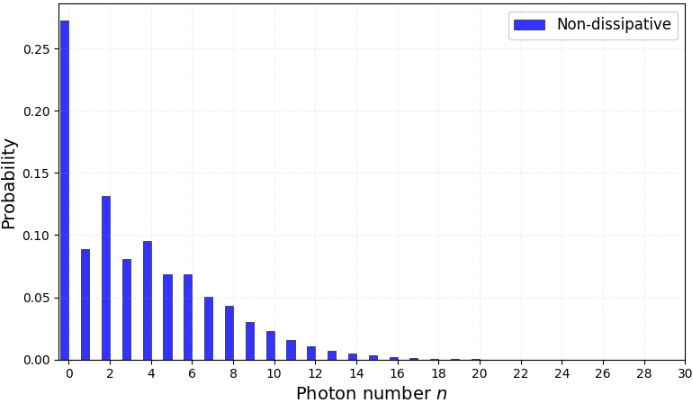}
    \caption{Photon number distribution \(P_1(n)\) for mode \(\hat{b}_1\) under non‑dissipative conditions $(\gamma_j=0$).}
    \label{fig:figPdisA1}
\end{figure}

\begin{figure}[htbp]
    \centering
    \includegraphics[width=1\linewidth]{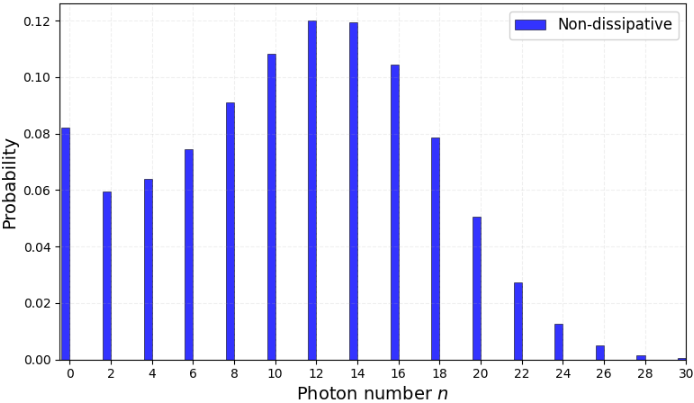}
    \caption{Photon number distribution \(P_3(n)\) for mode \(\hat{b}_3\) under non‑dissipative conditions $(\gamma_j=0)$.}
    \label{fig:figPdisA3}
\end{figure}

The photon number distribution serves as a diagnostic tool for identifying the parity pattern of an SCLS, revealing whether it corresponds to an even coherent state, an odd coherent state, or a mixture of even and odd components. This distribution is calculated via
\begin{equation}
P_j(n) = \operatorname{Tr}\bigl[ \hat{\rho}_j \, |n\rangle\langle n| \bigr].
\label{eq:pnd}
\end{equation}
Figs.~\ref{fig:figPdisA1}-\ref{fig:figPdisA3} display the photon number distributions (Eq.~\ref{eq:pnd}) for the SCLSs \(\hat{\rho}_j\). Under dissipation-free conditions, the distribution for mode \(\hat{b}_1\) corresponds to a mixture of even and odd coherent states, while that for mode \(\hat{b}_3\) corresponds to an even coherent state.


\subsection{Quadrature marginal distributions from the Wigner function}

From the Wigner function one obtains the probability densities of the two orthogonal quadratures $\hat{x}_j$ and $\hat{p}_j$ by integrating over the conjugate variable:

\begin{align}
	P(x_j) &= \int_{-\infty}^{\infty} W_j(x_j, p_j) \, dp_j, \\
	P(p_j) &= \int_{-\infty}^{\infty} W_j(x_j, p_j) \, dx_j .
\end{align}

These marginals are the measured distributions for homodyne detection with the local oscillator phase set to $0$ (for $P(x_j)$) or $\pi/2$ (for $P(p_j)$).

\begin{figure}[htbp]
    \centering
    \includegraphics[width=\linewidth]{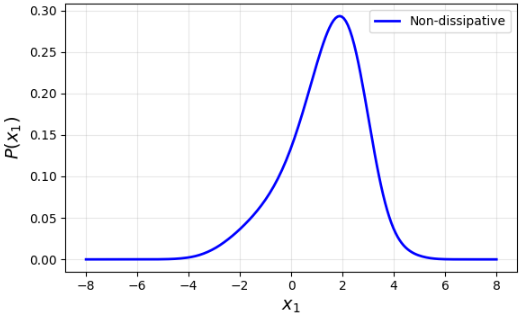}
    \caption{Probability distribution \(P(x_1)\) for mode \(\hat{b}_1\) in the non‑dissipative conditions $(\gamma_j=0$) at \(\tau = 0.190\), when the mean photon number \(n_3\) reaches its maximum.}
    \label{fig:figEndX1}
\end{figure}

\begin{figure}[htbp]
    \centering
    \includegraphics[width=\linewidth]{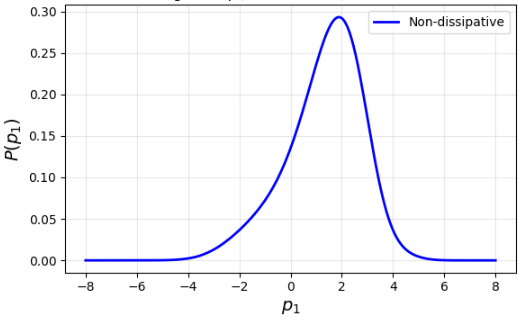}
    \caption{Probability distribution \(P(p_1)\) for mode \(\hat{b}_1\) in the non‑dissipative conditions $(\gamma_j=0$) at \(\tau = 0.190\), when the mean photon number \(n_3\) reaches its maximum.}
    \label{fig:figEndP1}
\end{figure}

\begin{figure}[htbp]
    \centering
    \includegraphics[width=\linewidth]{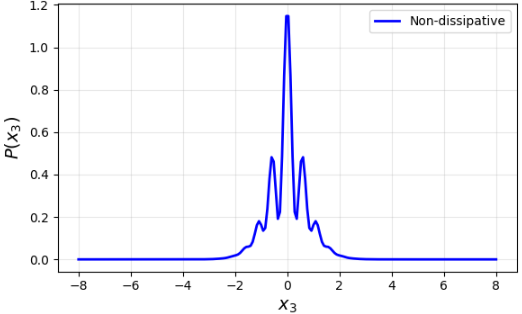}
    \caption{Probability distribution \(P(x_3)\) for mode \(\hat{b}_3\) in the non‑dissipative conditions $(\gamma_j=0$)  at \(\tau = 0.190\), when the mean photon number \(n_3\) reaches its maximum.}
    \label{fig:figEndX3}
\end{figure}

\begin{figure}[htbp]
    \centering
    \includegraphics[width=\linewidth]{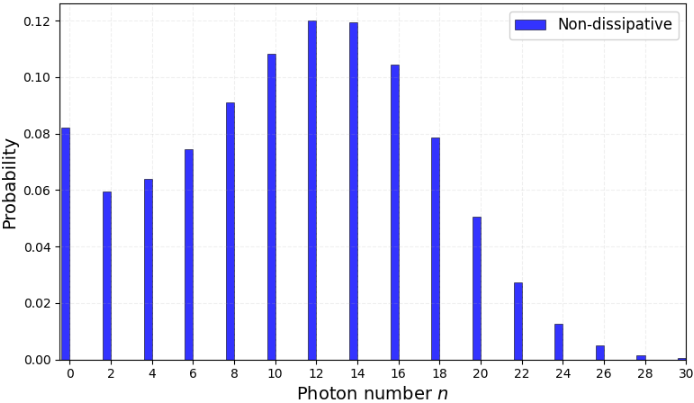}
    \caption{Probability distribution \(P(p_3)\) for mode \(\hat{b}_3\) in the non‑dissipative conditions $(\gamma_j=0$) at \(\tau = 0.190\), when the mean photon number \(n_3\) reaches its maximum.}
    \label{fig:figEndP3}
\end{figure}

The probability distributions (see Figs.~\ref{fig:figEndX1}-\ref{fig:figEndP3}) \(P(x_3)\) for mode \(\hat{b}_3\) in the non‑dissipative case show a structured profile (e.g., values rising from 0.02 to 1.95), which is characteristic of a SCLS with two well‑separated peaks in phase space.

\section{Results under the dynamics of $\hat{H}_{\text{int1}}$}

In this section, we present numerical results obtained from the original Hamiltonian \(\hat{H}_{\text{int1}}\) (Eq.~\ref{eq:hamiltonian}) without applying the transformation $\hat{a}_j(t) = e^{-i g \hat{N} t}  \, \hat{b}_j(t)$. The same parameter values are used as in the case with $\hat{H}_{\text{int2}}$. However, now the SPM and XPM terms are active. Consequently, the dynamics differ from the $\hat{H}_{\text{int2}}$ case.

\subsection{Wigner functions under $\hat{H}_{int1}$}
\begin{figure}[htbp]
    \centering    \includegraphics[width=1\linewidth]{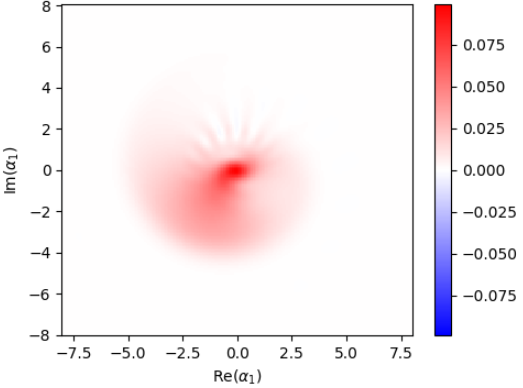}
    \caption{Wigner function for mode \(\hat{a}_1\) (non‑dissipative) using \(\hat{H}_{\text{int1}}\).}
    \label{fig:fig1ndfull}
\end{figure}

\begin{figure}[htbp]
    \centering    \includegraphics[width=1\linewidth]{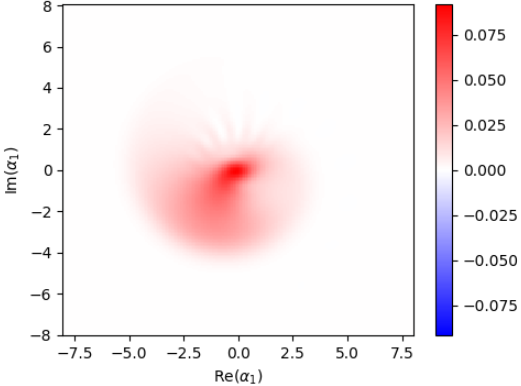}
    \caption{Wigner function for mode \(\hat{a}_1\) (dissipative) using \(\hat{H}_{\text{int1}}\).}
    \label{fig:fig1dfull}
\end{figure}

\begin{figure}[htbp]
    \centering    \includegraphics[width=1\linewidth]{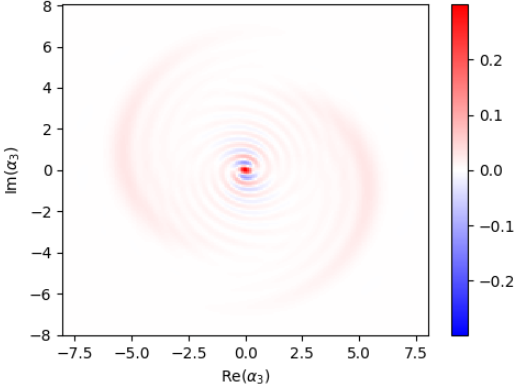}
    \caption{Wigner function for mode \(\hat{a}_3\) (non-dissipative) using \(\hat{H}_{\text{int1}}\).}
    \label{fig:fig3ndfull}
\end{figure}

\begin{figure}[htbp]
    \centering    \includegraphics[width=1\linewidth]{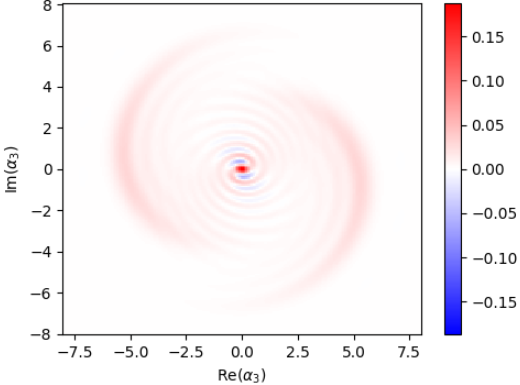}
    \caption{Wigner function for mode \(\hat{a}_3\) (dissipative) using \(\hat{H}_{\text{int1}}\).}
    \label{fig:fig3dfull}
\end{figure}

Figs.~\ref{fig:fig1ndfull}--\ref{fig:fig3dfull} show the Wigner functions at the time \(\tau_{\text{max}} = 0.190\) (the time where the mean photon number of mode \(\hat{a}_3\) reaches its maximum). In contrast to the decoupled case \(\hat{H}_{\text{int2}}\), the pump mode \(\hat{a}_1\) now exhibits a distorted Gaussian shape, while the signal mode \(\hat{a}_3\) still displays non‑Gaussian features. Nevertheless, the signal mode phase-space portraits (see Figs.~\ref{fig:fig3ndfull} and~\ref{fig:fig3dfull}) deviate from the canonical \cite{Dodonov1974} form expected for an even coherent state.

% (Include your actual figure files)
% (Add similar figures for dissipative, a3, etc.)

% (Include your actual figure files)
% (Add similar figures for dissipative, a3, etc.)

% (Include your actual figure files)
% (Add similar figures for dissipative, a3, etc.)
% (Include your actual figure files)
% (Add similar figures for dissipative, a3, etc.)
\subsection{Mean photon numbers and quadrature variances under $\hat{H}_{int1}$}
\begin{figure}[htbp]
    \centering   \includegraphics[width=\linewidth]{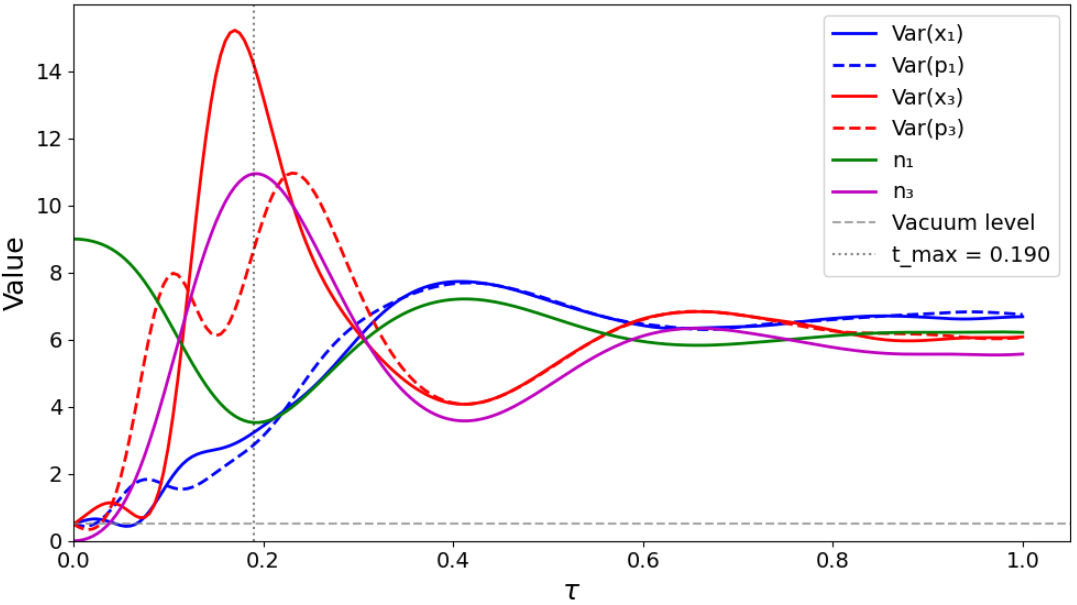}
   \caption{Evolution of mean photon numbers and quadrature variances for \(\hat{H}_{\text{int1}}\) (non‑dissipative).}
   \label{fig:fig1full}
\end{figure}

Fig.~\ref{fig:fig1full} displays the evolution of \(n_1\), \(n_3\) and the quadrature variances for the non‑dissipative case. At the extremal point \(\tau = 0.190\), the mean photon numbers and variances take the following values:
\begin{align*}
n_1 = 3.527376, \quad & n_3 = 10.9452,\\
\operatorname{Var}(x_1)=3.232426,\quad & \operatorname{Var}(p_1)=2.864985,\\
\operatorname{Var}(x_3)=14.199947,\quad & \operatorname{Var}(p_3)= 8.689676,
\end{align*}
indicating no squeezing in both modes at $\tau=0.190$.

\subsection{Entanglement under $\hat{H}_{\text{int1}}$}
The Schmidt number (Eq.~\ref{eq:ent}) evaluated at \(\tau = 0.190\) becomes \(K = 6.8582\), which is higher than the decoupled Hamiltonian $\hat{H}_{\text{int2}}$, showing that SPM and XPM increase the bipartite entanglement between the pumps and the signal.

\subsection{Fano factor under $\hat{H}_{int1}$}
The Fano factors at the extremal point \(\tau = 0.190\) are
\[
\begin{aligned}
\mathrm{FF}_1 &= 3.5072\;(\text{non‑diss.}),\; 3.3486\;(\text{diss.}),\\
\mathrm{FF}_3 &= 3.6269\;(\text{non‑diss.}),\; 3.5557\;(\text{diss.}),
\end{aligned}
\]
still above unity (super‑Poissonian) and similar to the case for $\hat{H}_{\text{int2}}$.

\subsection{Photon number distributions under $\hat{H}_{\text{int1}}$}

\begin{figure}[htbp]
    \centering
    \includegraphics[width=1\linewidth]{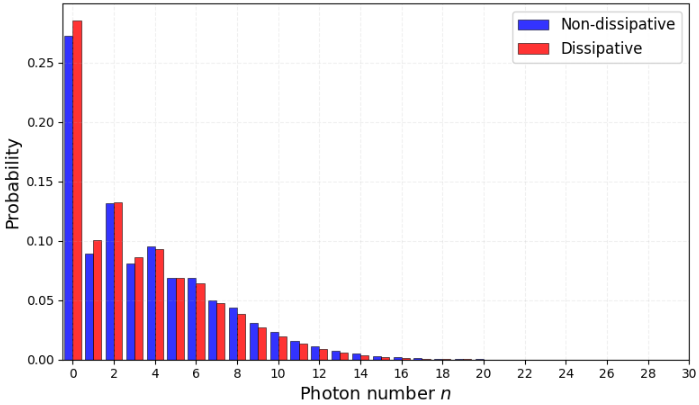}
    \caption{Photon number distribution for mode \(\hat{a}_1\) under \(\hat{H}_{\text{int1}}\).}
    \label{fig:figPdisA1full}
\end{figure}

\begin{figure}[htbp]
    \centering
    \includegraphics[width=1\linewidth]{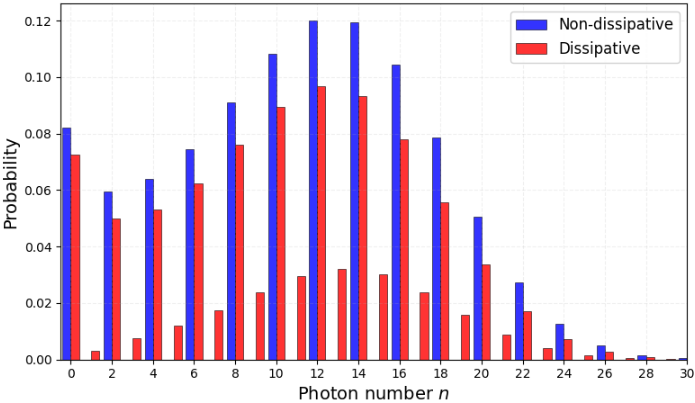}
    \caption{Photon number distribution for mode \(\hat{a}_3\) under \(\hat{H}_{\text{int1}}\).}
    \label{fig:figPdisA3full}
\end{figure}

Figure~\ref{fig:figPdisA1full} shows the pump mode $\hat{a}_1$ distribution.
For the signal mode $\hat{a}_3$ (Fig.~\ref{fig:figPdisA3full}), there is even‑photon dominance in the non‑dissipative case, while dissipation introduces odd photons in the dissipative case.


\subsection{Quadrature marginals under $\hat{H}_{\text{int1}}$}

\begin{figure}[htbp]
    \centering  \includegraphics[width=1\linewidth]{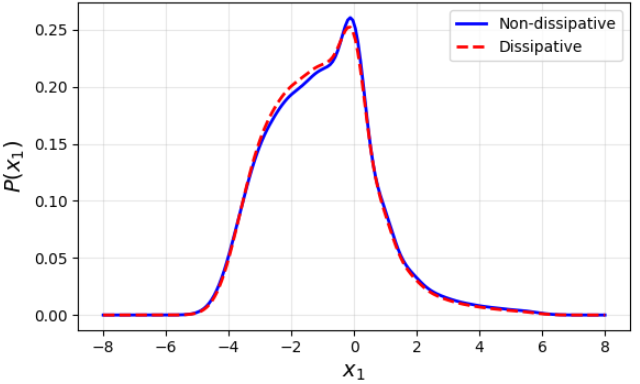}
    \caption{The quadrature marginal distribution \(P(x_1)\) for mode \(\hat{a}_1\) under \(\hat{H}_{\text{int1}}\) is shown.}
    \label{fig:figEndX1full}
\end{figure}

\begin{figure}[htbp]
    \centering  \includegraphics[width=1\linewidth]{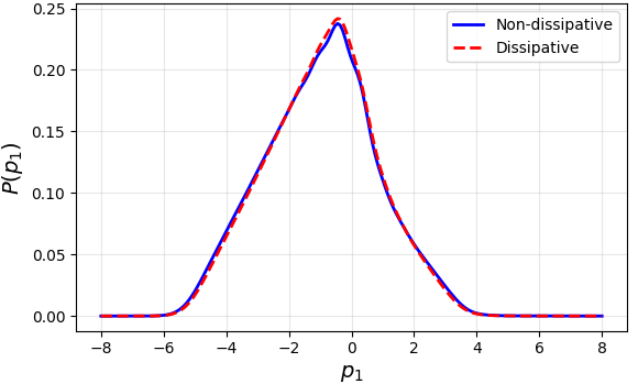}
    \caption{The quadrature marginal distribution \(P(p_1)\) for mode \(\hat{a}_1\) under \(\hat{H}_{\text{int1}}\) is shown.
}
    \label{fig:figEndP1full}
\end{figure}

\begin{figure}[htbp]
    \centering  \includegraphics[width=1\linewidth]{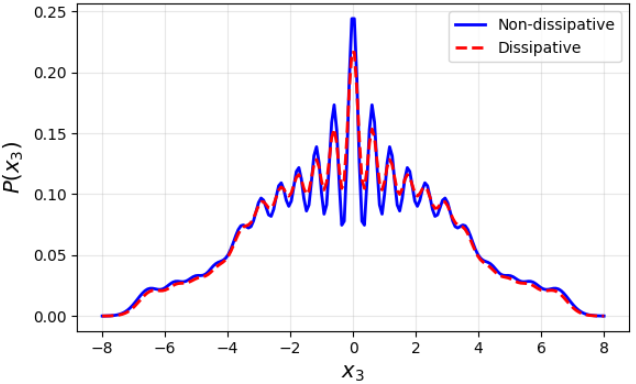}
    \caption{The quadrature marginal distribution \(P(x_3)\) for mode \(\hat{a}_3\) under \(\hat{H}_{\text{int1}}\) is shown.
}
    \label{fig:figEndX3full}
\end{figure}

\begin{figure}[htbp]
    \centering  \includegraphics[width=1\linewidth]{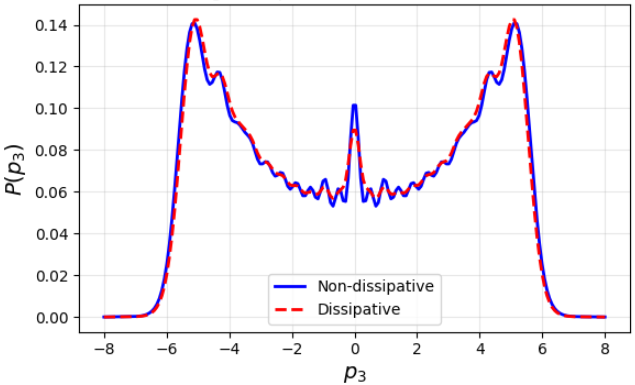}
    \caption{The quadrature marginal distribution \(P(p_3)\) for mode \(\hat{a}_3\) under \(\hat{H}_{\text{int1}}\) is shown.
}
    \label{fig:figEndP3full}
\end{figure}

Figures~\ref{fig:figEndX1full}--\ref{fig:figEndP3full} show the quadrature marginals \(P(x_1)\), \(P(p_1)\), \(P(x_3)\), and \(P(p_3)\) at $\tau=0.190$. We observe a negative impact on the interference pattern of the SCLS in the signal mode \(\hat{a}_3\), manifested as a reduction of the peak values at the center.

\subsection{Fidelity between non-dissipative and dissipative states of signal mode \(\hat{b}_3\) under $\hat{H}_{\text{int1}}$}

To quantify the similarity between the quantum states generated in the signal mode \(\hat{a}_3\) under non‑dissipative (\(\gamma_j = 0\)) and dissipative (\(\gamma_j = 0.2\)) conditions, we compute the fidelity. The non‑dissipative state \(\hat{\rho}_3^{(\mathrm{nd})}\) is pure (since there are no losses), while the dissipative state \(\hat{\rho}_3^{(\mathrm{d})}\) is mixed due to decoherence. For two density matrices \(\hat{\rho}\) and \(\hat{\sigma}\), the Uhlmann fidelity \cite{Jozsa1994} is defined as
\begin{equation}
    \mathcal{F}(\hat{\rho}, \hat{\sigma}) = \left( \Tr \left[ \sqrt{\sqrt{\hat{\rho}} \, \hat{\sigma} \sqrt{\hat{\rho}}} \right] \right)^2.
    \label{eq:fidelity_general}
\end{equation}
When one of the states is pure, e.g., \(\hat{\rho} = |\psi\rangle\langle\psi|\), the expression simplifies to \(\mathcal{F} = \langle\psi|\hat{\sigma}|\psi\rangle\). Using the appropriate form for our mixed‑state comparison, we evaluate the fidelity between the non‑dissipative and dissipative states at the extremal point \(\tau_{\text{max}} = 0.190\):
\begin{equation}
    \mathcal{F}\bigl(\hat{\rho}_3^{(\mathrm{nd})}, \hat{\rho}_3^{(\mathrm{d})}\bigr) = 0.903425.
\end{equation}
This value, close to unity, indicates that the dissipative state remains largely similar to the ideal non‑dissipative Schrödinger cat‑like state, despite the presence of decoherence.

\subsection{Summary of key differences}
Compared to the decoupled FWM model \((\hat{H}_{\text{int2}})\), the full or coupled Hamiltonian \((\hat{H}_{\text{int1}})\) leads to:
\begin{itemize}
    \item Creation of the SCLS in the signal mode, while the even-photon-number distribution remains intact.
    \item Schmidt number: Under \(\hat{H}_{\text{int2}}\): \(K = 2.02\); under \(\hat{H}_{\text{int1}}\): \(K = 6.86\) (stronger entanglement).
    
    \item Under \(\hat{H}_{\text{int1}}\), the fidelities between the dissipative and non‑dissipative states for both the pump and signal modes exceed \(0.9\).
\end{itemize}

\section{Results}
Semiclassical and parametric approximations did not capture the non-Gaussian features arising from pump depletion. In contrast, our full quantum treatment with a quartic interaction Hamiltonian -- without treating the pump as a classical undepleted field -- captures energy exchange between modes, phase‑dependent evolution, and quantum interference \cite{SinghTeretenkov2026,SinghBarinovAmosovMasalov2026}. Quantizing the pump mode fully accounts for its back‑action on the down‑converted fields, leading to non‑Gaussian Wigner functions with negative regions and interference fringes. As a direct consequence of a quantum‑mechanical pump description, SCLSs emerge in the signal mode. Operating in a high pump conversion regime, where the maximum number of photons from the pump modes is converted into the signal mode at \(\tau = 0.190\), we observe the formation of an SCLS — a regime essential for generating genuine non‑Gaussian states.

Unlike semiclassical or parametric approximations, our full quantum treatment with pump depletion produces non‑Gaussian features such as interference patterns and Wigner negativity, which are otherwise absent. Here, we study these previously overlooked non‑Gaussian properties in detail. 

Our results establish degenerate dual-pump spontaneous four-wave mixing in \(\chi^{(3)}\) microring resonators as a promising and robust platform for on‑chip generation of non‑Gaussian quantum states, specifically SCLSs. The demonstrated resilience to dissipation, along with the presence of entanglement and super‑Poissonian statistics, makes these states suitable for continuous‑variable quantum information processing \cite{Lloyd1999,Holevo2019} and quantum sensing \cite{Shukla2023,SinghTeretenkov2024}.

\section{ACKNOWLEDGMENTS}
R.S. is grateful to Prof. A.V. Masalov for insightful discussions.

\bibliographystyle{apsrev4-2}
\bibliography{bib} % matches your .bib filename without extension
\end{document}